\documentclass[pre,aps,preprint,showpacs]{revtex4}
\usepackage{revsymb,latexsym,amssymb,amsmath,comment}
\usepackage{graphicx,psfrag,subfigure,stmaryrd}

\begin{document}
\author{Th. Michael, S. Trimper}
\affiliation{Institut f\"ur Physik,
Martin-Luther-Universit\"at, D-06099 Halle Germany}
\email{thomas.michael@physik.uni-halle.de;steffen.trimper@physik.uni-halle.de}
\author{J. M. Wesselinowa }
\affiliation{University of Sofia, Department of Physics\\ Blvd. J. Bouchier 5, 
1164 Sofia, Bulgaria}
\email{julia@phys.uni-sofia.bg}
\title{Size and doping effects on the coercive field of ferroelectric nanoparticles} 
\date{\today }
\begin{abstract}
A microscopic model for describing ferroelectric nanoparticles is proposed which allows us to 
calculate the polarization as a function of an external electric field, the temperature, the defect 
concentration and the particle size. The interaction of the constituents of the material, 
arranged in layers, depends on both the coupling strength at the surface and that of defect shells 
in addition to the bulk values. The analysis is based on an Ising model in a transverse field, 
modified in such a manner to study the influence of size and doping effects on the hysteresis loop 
of the nanoparticles. Using a Green's function technique in real space we find the coercive field, 
the remanent polarization and the critical temperature which differ significantly from the bulk behavior. 
Depending on the varying coupling strength due to the kind of doping ions and the surface configuration, the 
coercive field and the remanent polarization can either increase or decrease in comparison to 
the bulk behavior. The theoretical results are compared with a variety of different experimental data.


\end{abstract}

\maketitle

\section{Introduction}

\noindent The interesting physical properties and the ability to store information via a switchable 
polarization have moved ferroelectric materials into the focus of research. One important application, 
the non-volatile memories with ferroelectric capacitor materials are known as ferroelectric random 
access memories (FRAM). The processing issues involved in the high density integration process are 
highly dependent on the ferroelectric and electrode-barrier materials. Hence, the selection of materials 
is a decisive factor in determining the performance of the device \cite{scott}. In view of fundamental 
ferroelectric properties, there are two potential ferroelectric materials for FRAM applications, namely 
Pb(Zr,Ti)TiO$_3$ (PZT) and SrBi$_2$Ta$_2$O$_9$ (SBT) \cite{1}. They possess a high remanent polarization 
$\sigma_r$, low coercive electric field $E_c$, which characterizes the polarization reversal, and low 
dielectric loss. The use of ferroelectric thin films and small particles in high density non-volatile 
random access memories is based on the ability of ferroelectrics being positioned in two opposite 
polarization states by an external electric field \cite{2}. It is therefore of great interest to study 
the thickness dependence of $E_c$ for small ferroelectric (FE) particles. This coercive electric field 
usually increases significantly with decreasing film thickness or particle size \cite{3,4,5,6}. The 
strength of the coercive field is related to the ease of domain nucleation and domain-wall motion, 
while the permittivity is related to the density of domain walls and their mobility at low fields. 
A diversity of different explanations have been proposed in the past for this size effect, including 
the surface pinning of domain walls and the influence of an internal electric field on the domain 
nucleation in depleted films. Current observations state that the thickness dependence of the coercive 
field in PZT films with metallic electrodes is caused mainly by the presence of a non-ferroelectric 
layer at the film/electrode interface. A parabolic-like relationship between the coercive field and 
grain size of ferroelectric PZT films is obtained by Liu et al. \cite{7}. Nevertheless, the curvature 
of the parabolic curves varies with the thickness. It is demonstrated that the curvature increases 
as the thickness of the film decreases. The influence of thickness on the curvature was explained 
by the dielectric dependence on the thickness. The coercive field of a FE particle is stress sensitive. 
Lebedev and Akedo \cite{8} have shown that the increase of internal compressive stress for thinner 
PZT films led to the increase of the coercive field and breakdown electrical strength, while tensile 
stress decreases $\sigma_r$ and $E_c$.

\noindent In order to obtain high remanent polarization and low coercive field there are many experiments with 
doped ferroelectric thin films and small particles. By adding oxide group softeners, hardeners and 
stabilizers one can modify these materials. Softeners (donors) reduce the coercive field strength 
and the elastic modulus and increase the permittivity, the dielectric constant and the mechanical losses. 
Doping of hardeners (acceptors) gives higher conductivity, reduces the dielectric constant and increases 
the mechanical quality factor \cite{9}. 

\noindent Multiple ion occupation of A and/or B sites in ABO$_3$ compounds can affect the lattice parameters 
and the tetragonal distortion ($c/a$). As a consequence a change of the physical properties as the 
polarization and the phase transition temperature is expected. Direct evidence of A-site-deficient 
SBT and its enhanced ferroelectric characteristics is given by Noguchi et al. \cite{10}. The increase of 
Ba concentration in PLZT ceramics done by Ramam \cite{14} and the substitution of La in PZT nanopowders 
and thin films lead to a marked decrease in $T_c$, $\sigma_r$ and $E_c$ \cite{11,12,13}. An enhancement 
of dielectric constant and lower $T_c$ and $E_c$ were observed by addition of PT particles to a PZT 
matrix \cite{15}.

\noindent There is only a small number of theoretical studies of the dependence of the coercive field 
in ferroelectric small particles. Lo \cite{16} has investigated theoretically the thickness effect in 
ferroelectric thin films using Landau-Khalatnikov theory. Ferroelectric properties such as the hysteresis 
loop, and its associated coercive field and the remanent polarization of various film thickness have been 
numerically simulated. Recently FE thin films has been studied in dependence on the film thickness, 
where the phenomenological Landau theory is applied \cite{wz}. The authors stress that the thickness-dependent 
coercive field represents the intrinsic coercive field. Moreover, they point out to the role of defects 
observed in real ferroelectric thin films. In general the number of defects increases with the film thickness.
 
\noindent The aim of the present paper is to study the size dependence and doping effects of ferroelectric 
small particles using a more refined model, in which the microscopic interaction between the constituent parts 
of the material are explicitly taken into account. A adequate candidate is the Ising model in a transverse 
field (TIM), where ferroelectric properties are described by pseudo-spin operators. The model is quite 
successful for bulk material \cite{blinc,cao}. Here we extend the applicability of the model to nanoparticles. 
We utilize a Green's function technique to investigate the interacting system, which has to be formulated in 
real space because of the lack of translational invariance.

\section{The model and the Green's function}

\begin{figure} [!ht]
\centering
\psfrag{si}[][][1.3]{${\sigma}$}
\psfrag{Temp}[][][0.75]{$T [K]$}
\includegraphics[scale=0.75]{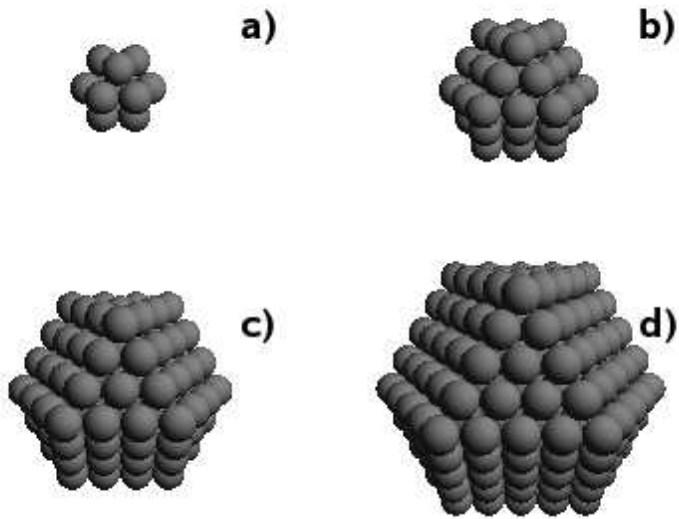}
\caption{Array of the ferroelectric nanoparticles composed of different shells. 
Each sphere represents a spin situated in the center, where (a) consists of one central spin plus 
$N=1$ shell,\\(b)\, $N=2$,\, (c)\,$N=3$,\,(d)\, $N=4$\,.}
\label{Fig.0}
\end{figure}

\noindent  Following Blinc and \u{Z}ek\u{s} \cite{blinc} the FE constituents are described by 
pseudo-spin operators. We consider a spherical particle characterized by fixing the origin at 
a certain spin in the center of the particle. All other spins within the particle are ordered 
in shells, which are numbered by $n=0,1,...,N$. Here $n=0$ denotes the central spin and $n=N$ 
represents the surface of the system, see Fig.~\ref{Fig.0}. 

\noindent The surface effects are included by different coupling parameters 
within the surface shell and within the bulk.\\
The Hamiltonian of the TIM includes both, bulk and surface properties. It reads
\begin{equation}
H=-\frac{1}{2}\sum_{ij}J_{ij}S^z_iS^z_j-\sum_{i}\Omega_iS^x_i-\mu E \sum_i S^z_i,
\label{ham}
\end{equation}
where $S_i^x$ and $S_i^z$ are components of spin-$\frac{1}{2}$ operators and the sums are performed over 
the internal and surface lattice points, respectively. $E$ is an external electrical field. The 
quantity $J_{ij}$ represents the strength of the interaction between spins at nearest neighbor sites 
$i$ and $j$. The interaction of spins at the surface shell is denoted as $J_{ij} = J_s$, whereas 
otherwise the interaction is $J_b$. In the same manner $\Omega_b$ and $\Omega_s$ represent transverse 
fields in the bulk and surface shell. It is important to mention that the exchange interaction 
$J_{ij} \equiv J(r_i-r_j)$ depends on the distance between the spins, consequently the interaction 
strength is determined by the lattice parameters, the lattice symmetry and the number of next 
nearest neighbors. Starting from Eq.~(\ref{ham}), compare \cite{blinc}, one finds immediately that 
the ordered phase is characterized by $<S^x>\, \neq 0$ and $<S^z>\, \neq 0$. Therefore it is appropriate 
to introduce a new coordinate system by rotating the original one by the angle $\theta$ in the $x-z$ plane. 
The rotation angle $\theta$ is determined by the requirement $<S^{x'}>\, = 0$ in the new coordinate system.

\noindent Since the lack of translational invariance the Green's function has to be investigated in the 
real space. The retarded Green's function is defined by
\begin{equation}
G_{lm}(t)=\ll{S^+_l(t);S^-_m(0)}\gg\,.
\label{gf}
\end{equation}
The operators $S_l^+, S_m^-$ are the Pauli operators in the rotated system. The equation of motion 
of the Green's function in random phase approximation (RPA) reads
\begin{eqnarray}
\omega G_{lm}&=& 2 \langle S^z_l \rangle \delta _{lm} +\Big[2\Omega_l \sin \theta_l+\mu E \cos\theta_l+ 
\sum_j {J_{lj}\cos\theta_l \cos\theta_j \langle S^z_j \rangle} \nonumber\\
&+&\frac{1}{2} \sum_jJ_{lj} \sin\theta_l \sin\theta_j(\langle S^+_l S^-_j \rangle +\langle S^-_l S^-_j \rangle)\Big]G_{lm}\nonumber\\
&-&\frac{1}{2}\sum_j {J_{lj}\left[\sin\theta_l \sin\theta_j \langle S^z_l \rangle+ 2\cos \theta_j \cos \theta_l 
\langle  S^+_l S^-_j \rangle \right] G_{jm}}.
\label{gf1}
\end{eqnarray}
The poles of the Green's function give the transverse excitation energies. Within the applied RPA 
the transverse spin-wave energy is found as
\begin{equation}
\epsilon _n = 2 \Omega _n \sin\theta _n + \frac{1}{N'}\sum_j J_{nj} \cos\theta _n \cos\theta _j 
\langle S_j^z\rangle + \mu E \cos\theta _n\,, 
\label{te}
\end{equation}
where $N'$ is the number of sites in any of the shells. To complete the soft-mode energy $\epsilon_n $ 
of the $n$-th shell, one needs the rotation angle $\theta_n$, which follows from the condition  
$<S^{x'}>\, = 0$. From here the angle is determined by the equation
\begin{equation}
-\Omega_n \cos\theta_n + \frac{1}{4} \sigma_n J_n \cos\theta_n \sin\theta_n + \frac{1}{2} \mu E \sin\theta_n =0.
\label{ang}
\end{equation}
Using the standard procedure for Green's function we get the relative polarization of the $n$-th shell 
in the form
\begin{equation}
\sigma_n= \langle S^z_n \rangle =\frac{1}{2}\tanh{\frac{\epsilon_n}{2T}}\,.
\label{pol}
\end{equation}

\section{Numerical Results and Discussion}

\begin{figure} [!ht]
\centering
\psfrag{si}[][][1.3]{${\sigma}$}
\psfrag{Temp}[][][0.75]{$T []$}
\includegraphics[scale=0.75]{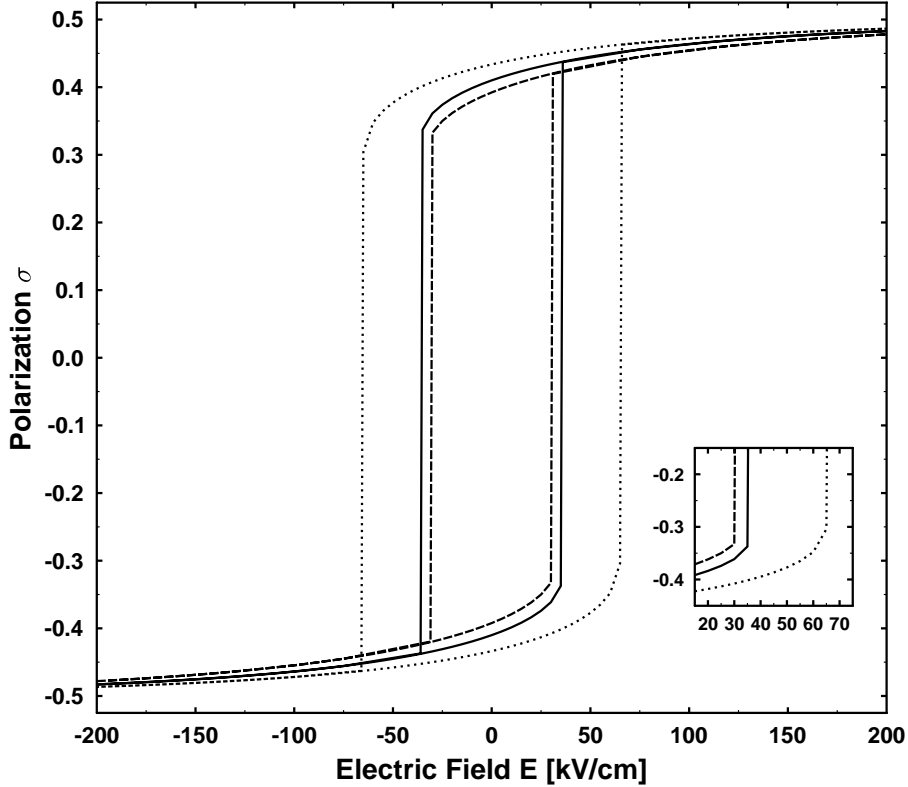}
\caption{Influence of the surface coupling strength $J_s$ on the hysteresis at fixed temperature  
$T = 300\,\rm{K}$,\, $J_b = 150\,\rm{K}$ and $N = 8$: $J_s = 150\,\rm{K}$ (solid curve), 
$J_s = 350\, \rm{K} > J_b $ (dotted curve), $J_s = 50\, \rm{K} < J_b$ (dashed curve); 
the insert offer the low field behavior\,.}
\label{Fig.1}
\end{figure}

\noindent In this section we present the numerical results of the relevant quantities based on 
Eqs.~(\ref{te}, \ref{ang}, \ref{pol} ). In particular, we analyze the behavior of spherical FE particles, 
which are depicted in Fig.~\ref{Fig.0}. 
In case of BTO the bulk interaction parameter is $J_b = 150\,$K and the transverse field is 
$\Omega_b= 10\,$K. Due to the different numbers of next nearest neighbors on the surface the 
interaction strength $J$ can take different values for the surface, denoted by $J_s$, and for the bulk,  
labeled as $J_b$, compare also \cite{17}. 
Firstly, let us consider the hysteresis loop for different surface configurations represented 
by the interaction constant $J_s$ at a fixed temperature $T= 300 \,$K and fixed $N$. The results 
for a particle with $N=8$ shells are shown in Fig.~\ref{Fig.1}. 

\noindent One observes that the coercive field $E_c$ is sensitive to variations of the 
interaction parameter $J$. If the coupling at the surface is smaller $J_s = 50 $K as that 
in the bulk $J_b = 150$ K (dashed line), the coercive field $E_c$  and the remanent polarization 
$\sigma_r$ are reduced in comparison to the case for $J_s = J_b$ (solid line). With other words, 
the coercive field is lowered when the critical temperature of the system is decreased due to the 
smaller $J_s$ value. In the opposite case $J_s > J_b$ (dotted line) both, the coercive field $E_c$ 
and the remanent polarization $\sigma_r$ increase compared to $J_s = J_b$. Because of the enlarged 
value of $J_s$ the phase transition temperature $T_c$ of the small particle will be enhanced. The 
realization $J_s < J_b$ may explain the decrease of the polarization $\sigma$ and the phase 
transition temperature $T_c$ observed in small BTO- \cite{18} and PTO- particles \cite{19}. The 
opposite case, namely a stronger coupling at the surface, is in agreement with observations made 
in small KDP- particles \cite{20}, where the polarization and the critical temperature increase 
compared to the bulk material.

\begin{figure} [!ht]
\centering
\psfrag{si}[][][1.3]{${\sigma}$}
\psfrag{Temp}[][][0.75]{$T [K]$}
\includegraphics[scale=0.75]{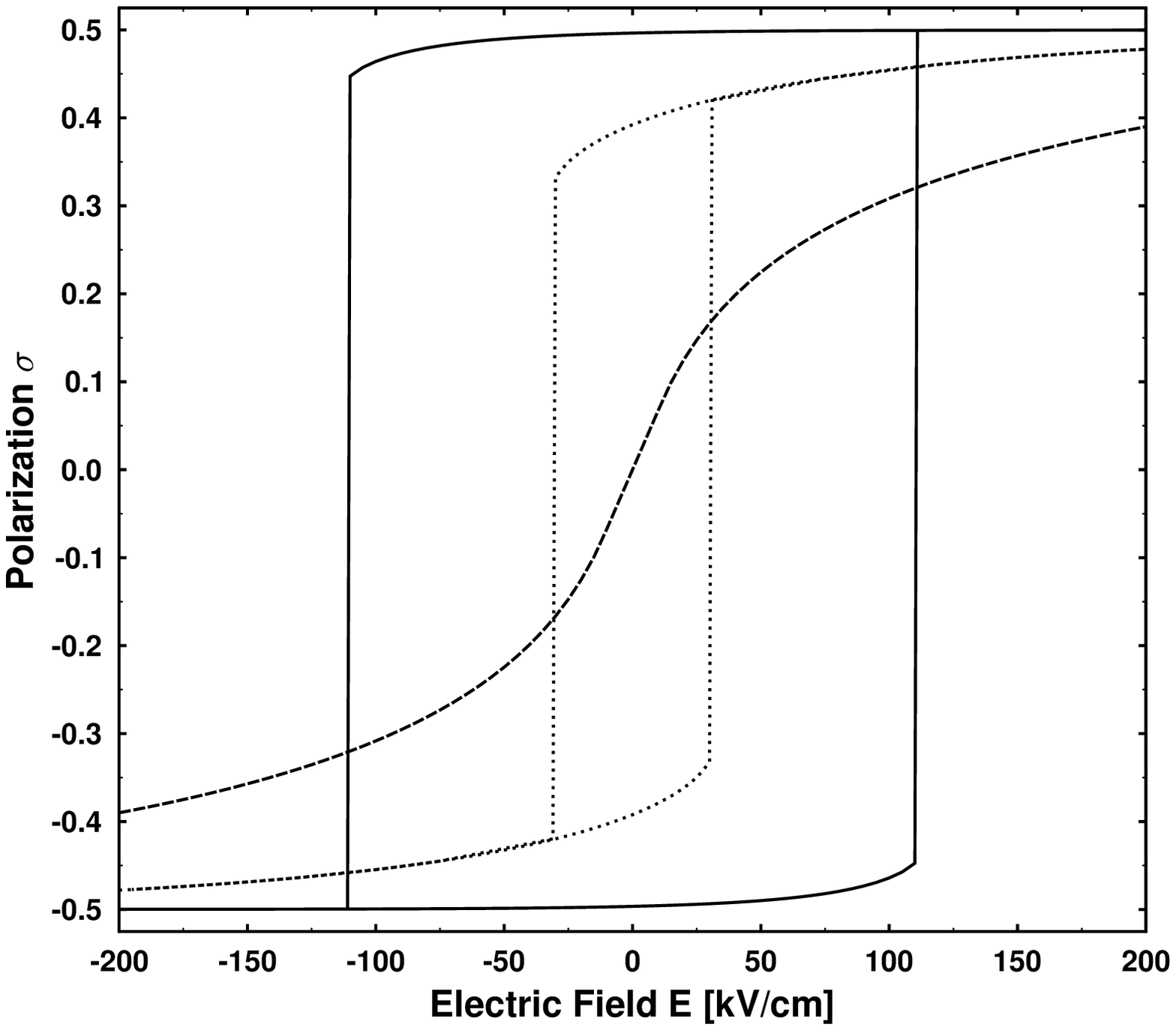}
\caption{Temperature dependence of the hysteresis with $N=8,\, J_b = 150\, \rm{K},\, J_s = 50\, \rm{K} $: 
$T = 100\, \rm{K}$ (solid curve), $T = 300\, \rm{K} $ (dotted curve ),\, $T = 500\, \rm{K}$ (dashed curve)\,.}
\label{Fig.2}
\end{figure}

\noindent In order to investigate the properties of BTO-type FE small particles we examine in 
the following the case $J_s < J_b$ in more detail. The temperature dependence of the hysteresis 
loop for eight layers is shown in Fig.~\ref{Fig.2}.

\begin{figure} [!ht]
\centering
\psfrag{si}[][][1.3]{${\sigma}$}
\psfrag{Temp}[][][0.75]{$T [K]$}
\includegraphics[scale=0.75]{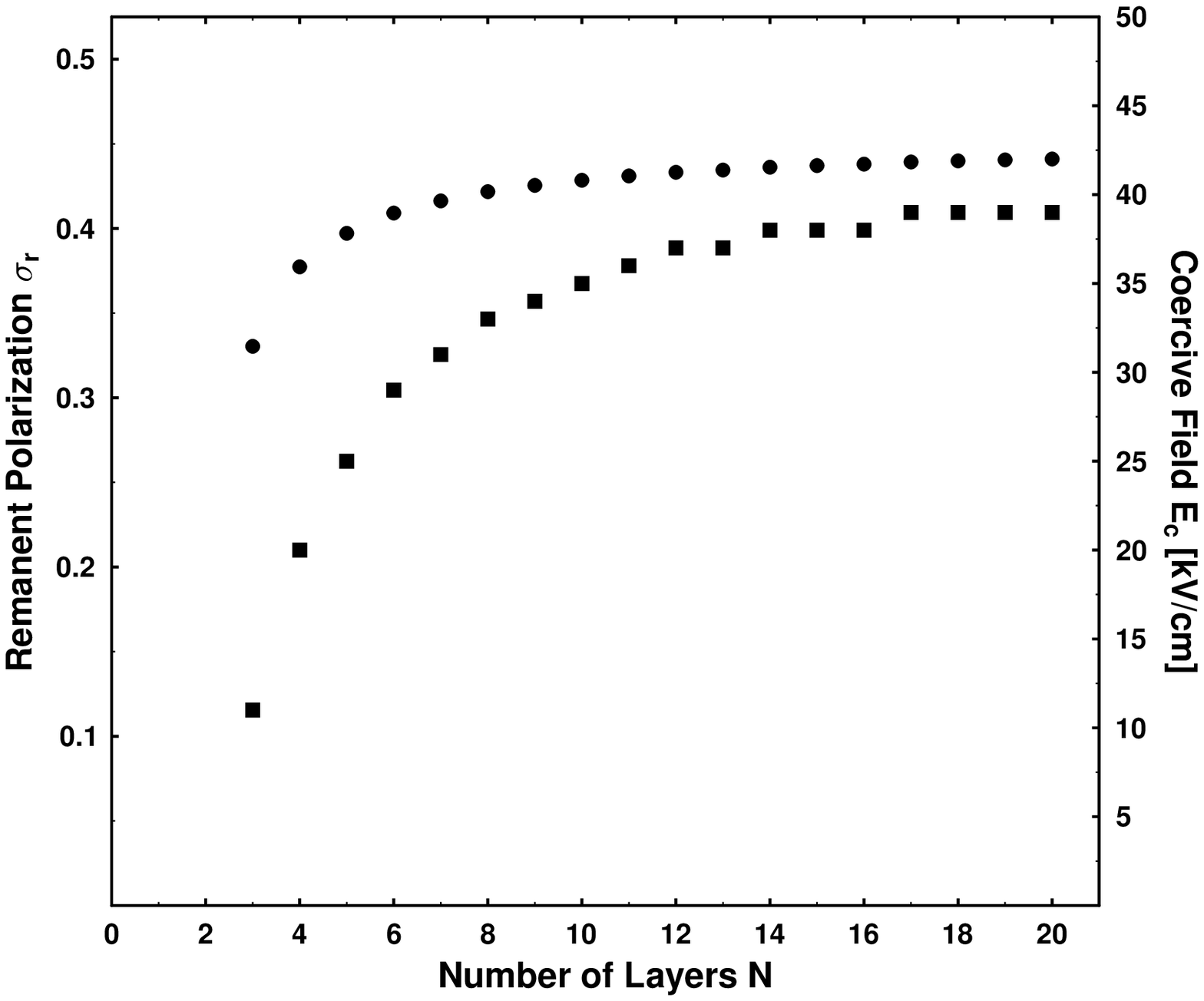}
\caption{Remanent polarization $\sigma_r$ (circles) and coercive field $E_c$ (squares) for different 
layers $N$ at temperature $T = 300\, \rm{K}$ and couplings $J_b = 150\, \rm{K},\, J_s = 50\, \rm{K} $\,.}
\label{Fig.3}
\end{figure}

\noindent With increasing temperature the hysteresis 
loop is more compact and lower, the coercive field decreases and even for $T \geq T_c$ the hysteresis 
loop vanishes (dashed line). Moreover, we have investigated the dependence of the remanent polarization 
$\sigma_r$ and the coercive field $E_c$ on the particle size $N$. The results are depicted in 
Fig.~\ref{Fig.3}. 

\noindent Both the field $E_c$ (square) and the polarization $\sigma_r$ (circle) increase with 
increasing $N$, which is in agreement with theoretically finding \cite{wz} and the experimental 
data from ultrathin FE BTO- films \cite{kkk}. We found in our approach no rising coercive field 
with decreasing particle size as reported in \cite{3,4,5,6}.

\begin{figure} [!ht]
\centering
\psfrag{si}[][][1.3]{${\sigma}$}
\psfrag{Temp}[][][0.75]{$T [K]$}
\includegraphics[scale=0.75]{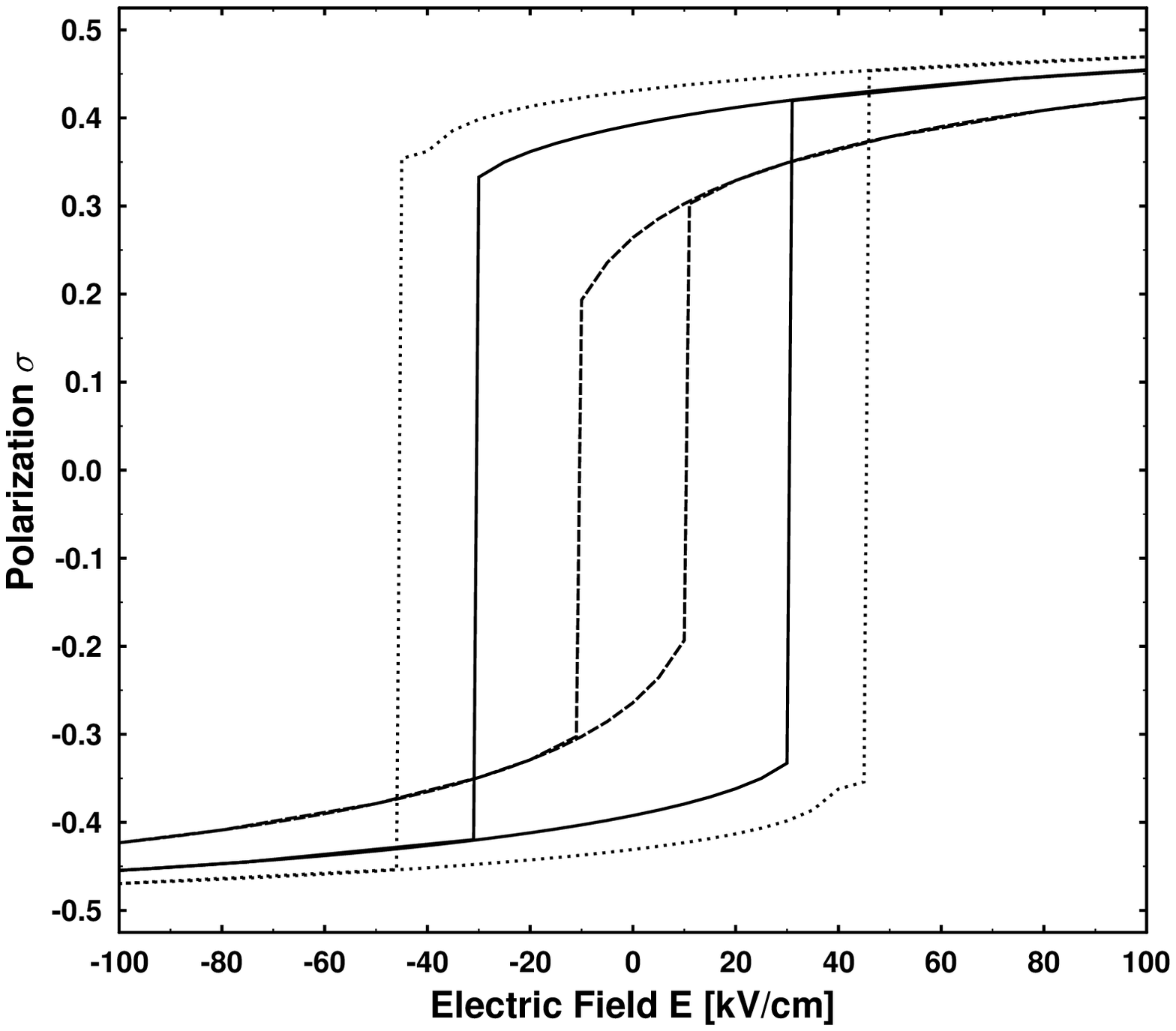}
\caption{Influence of defects on the hysteresis for $T = 300\, \rm{K},\, J_b = 150\, \rm{K},\, 
J_s = 50\, \rm{K}$\,. From the total number of $N=8$ layers the first five layers are defect layers: 
$J_d = J_b $ (solid curve); $J_d = 400\, \rm{K} $ (dotted curve);$J_d = 25\, \rm{K}$ (dashed curve).}
\label{Fig.4}
\end{figure}

\noindent Because there are several experimental indications for a significant influence of doping 
effects on the hysteresis loop, now our model is modified in such a manner to include defects. Physically, 
such defects can be originated by localized vacancies or impurities, doping ions with smaller radii 
and larger distances in-between in comparison to the host material, see also \cite{wz}. Microscopically,   
the substitution of defects into the material leads to a change of the coupling parameter. Within our 
model let us assume that one or more of the shells are composed of defects. The interaction strength  
between neighbors $J_d$ is altered and in general different from the surface value $J_s$ as well as 
the bulk one $J_b$. Furthermore, the defect can be situated at different shells within the nanoparticle. 
In Fig.~\ref{Fig.4} we show the influence of defects with variable strength $J_d$ on the hysteresis loop. 

\noindent A single isolated defect layer offers only a weak influence on the hysteresis curve. 
Due to that we have studied the case that the first five layers of the nanoparticle are defect 
ones (compare also Fig.~\ref{Fig.0}). Here the number of 
ferroelectric constituents is large enough to give a significant contribution to the polarization and 
consequently to the hysteresis loop. Apparently one observes a change of the shape of the hysteresis 
loop due to defects. In real materials one expect $J_d < J_b$. In that case the coercive field and 
the remanent polarization 
are reduced (dashed curve) compared to the case without defects $J_d = J_b$ (solid curve). This may 
explain the experimentally observed decrease of $T_c$, $E_c$ and $\sigma_r$ in small FE 
particles by the substitution of doping ions. The situation is for instance realized by substituting 
La in PTO \cite{11} and PZT \cite{12,13} nanopowders. Although in the majority of real materials $J_d$ 
should be smaller than $J_b$, we study also the opposite case $J_d > J_b$. The results are shown as 
the dotted curve in Fig.~\ref{Fig.4}. The coercive field, the remanent polarization and the critical 
temperature are larger than those without defects (solid curve in (Fig.~\ref{Fig.4})). Such a situation, 
namely that the defect coupling is stronger than the bulk coupling will be realized, when the impurities 
have a larger radius compared to the constituent ions. The corresponding quantities are enhanced in 
comparison to the bulk value, which is clearly originated by an enhanced $J_d$-coupling, which is 
in accordance with the experimentally observed increase of $T_c$, $E_c$ and $\sigma_r$ by 
the substitution of doping ions, such as Bi in SBT \cite{10} or by increasing the Ba contents in 
PLZT ceramics \cite{14}. This behavior is in contrast to the case $J_d < J_b$. Let us stress that 
the variation of the interaction strength $J$ can be also interpreted as the appearance of local 
stress, originated by the inclusion of different 
kinds of defects. The case $J_d > J_b$ corresponds to a compressive stress, leading to an enhancement 
of $E_c$, which has been observed in thin PZT films \cite{8}. The opposite case, i.e. tensile stress, 
yields to a decrease of $E_c$. 

\begin{figure} [!ht]
\centering
\psfrag{si}[][][1.3]{${\sigma}$}
\psfrag{Temp}[][][0.75]{$T [K]$}
\includegraphics[scale=0.75]{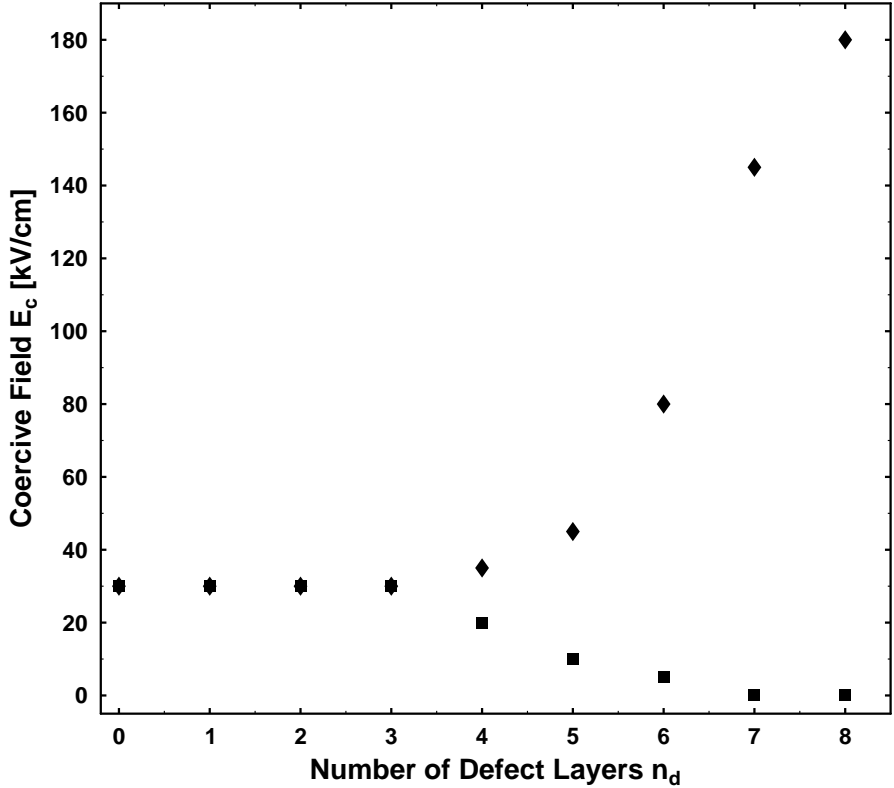}
\caption{Coercive field $E_c$ in dependence on the number of defect layers $n_d$ for 
$N = 8,\,J_b = 150\, \rm{K},\, J_s = 50\, \rm{K}, T = 300\, \rm{K}$ and different strength 
of $J_d$: $J_d = 25\, \rm{K}$, i. e. $J_b > J_d$ (squares) and $J_d = 400\, \rm{K}$, i. e. 
$J_d > J_b $ (diamonds) \,. }
\label{Fig.5}
\end{figure}

\noindent As visible in Figs.~(\ref{Fig.3},\,\ref{Fig.4}) the coercive field and  
the remanent polarization of the FE particle are decreased or increased due to the different interaction 
strength within the defect shell. Obviously the mentioned 
quantities should depend on the number of the inner defect shells, i. e. on the 
concentration of the defects. This dependence is shown in Fig.~\ref{Fig.5} for a particle with eight shells.

\noindent Notice that for instance $n_d = 5$ means, that all layers 
until the fifth layers are defect layers, a situation which is precisely considered in Fig.~\ref{Fig.4}. 
The squares in Fig.~\ref{Fig.5} demonstrates, that the coercive field strength 
$E_c$ decreases with increasing number of defect shells, where we assume $J_d = 25 \,$K, i.e. $J_b > J_d$. 
This result is in reasonable accordance to the experimental data given in \cite{11,12,13}. A similar result 
is also obtained for the remanent polarization $\sigma_r$. An increase of the La content in PTO and PZT ceramics 
decreases the coercive field $E_c$. The opposite behavior is offered as the diamonds in 
Fig.~\ref{Fig.5} with $J_d = 400\,$K, i.e. $J_d > J_b $. With increasing number of defect layers 
the coercive field $E_c$ (respectively $\sigma_r$) increases. This finding is in a quite good agreement 
with the experimental data offered by Noguchi et al. \cite{10} and Ramam et al. \cite{14}.

\begin{figure} [!ht]
\centering
\psfrag{si}[][][1.3]{${\sigma}$}
\psfrag{Temp}[][][0.75]{$T [K]$}
\includegraphics[scale=0.75]{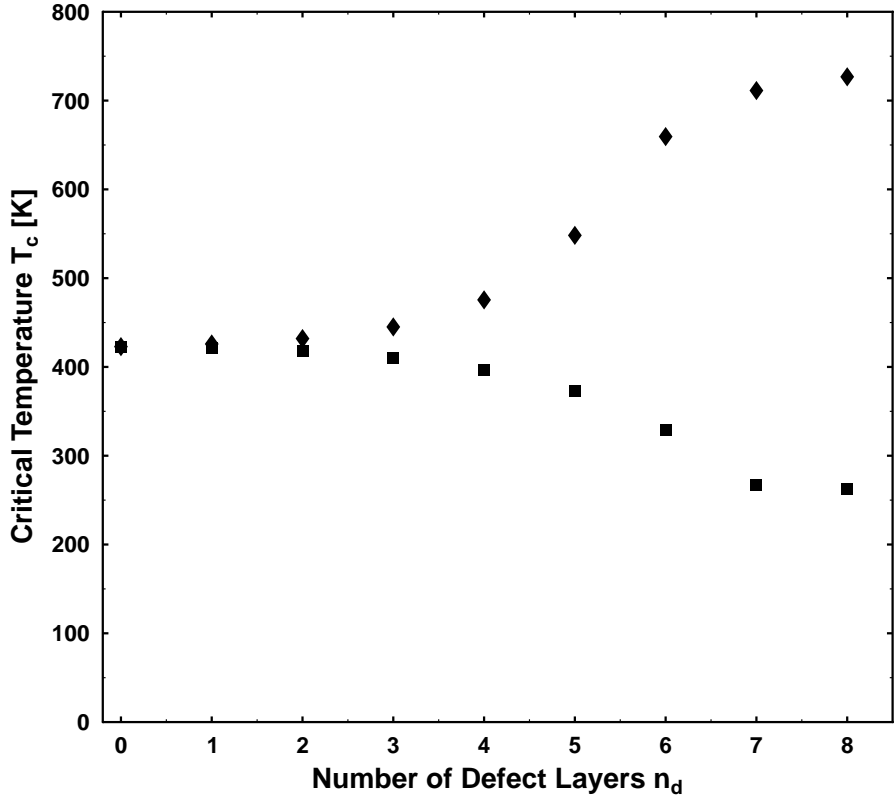}
\caption{Dependence of the critical temperature $T_c$ on the number of defect layers $n_d$ for 
$N = 8,\,J_b = 150\, \rm{K},\, J_s = 50\, \rm{K}$ and different $J_d$ couplings: 
$J_d = 25\, \rm{K}$, i. e. $J_b > J_d$ (squares) and $J_d = 400\, \rm{K}$, i. e. $J_b < J_d$ (diamonds)\, .}
\label{Fig.7}
\end{figure}

\noindent Fig.~\ref{Fig.7} shows the dependence of the Curie temperature $T_c$ on the number 
of defect layers $n_d$ for a spherical particle. 

\noindent A defect coupling smaller than the bulk and surface couplings, $J_d < J_b$ (squares), yields a 
decreasing critical temperature $T_c$  with increasing number of defect shells, whereas in the opposite 
case (diamonds) the critical temperature increases with the number of defect layers. The different 
behavior is controlled by the interaction strength. 
These results are in a qualitative agreement with the experimental data obtained by Noguchi et al. \cite{21}. 
They have studied SBT films ($T_c = 295\,^0C$) where the rare earth cations have been substituted by 
La, Ce, Pr, Nd and Sm as well as Bi at the A site (Sr site) with Sr vacancies. The La modification 
induces a soft behavior (lowering of $E_c$ and $P _r$), while a large amount of Nd and Sm substitution 
results in a very high 
$E_c$ (hard). So the modified behavior is the result of defect engineering of both Sr
and oxide vacancies. Likewise other bismuth layer-structured FE (BLSF) reveals that 
$T_c$ is strongly influenced by the radius $r_i$ of A-site cations. BLSF with smaller A-site cations 
(Ca$^{2+}$) tends to a higher $T_c$ (420\,$^0$C). The inclusion of the same amount of Ba$^{2+}$ (with larger 
radius) leads to a relaxation of FE distortions and results in a decrease of $T_c$ to
120\,$^0$C. The substitution of La with concentration $x = 0.5$ gives rise to a marked decrease 
in $T_c$ to 180\,$^0$C, because the induced A-site vacancies weaken the coupling between neighboring 
BO$_6$ octahedral \cite{21,22}. This result corresponds in our approach when the interaction parameter 
$J_d$ is reduced, i. e. $J_d < J_b$. In La-modified PbTiO$_3$ the transition temperature $T_c$ 
decreases significantly, too, with an increase of the La content \cite{23}. In Bi-SBT 
the critical temperature $T_c$ increases strongly to $405\,^0$C ($x$=0.2) \cite{21}. 
The increase in $T_c$ by Bi substitution tends to the opposite behavior as that observed in La-SBT. This 
is governed by the bonding characteristics with oxide ions. The influence of the orbital 
hybridization on $T_c$ is very large, and Bi substitution results in a higher $T_c$ \cite{24}.
This experimental observation is in a sufficient good agreement with our analytical finding when the 
interaction strength satisfies $J_d > J_b$.

\section{Conclusion}

\noindent Because ferroelectricity at a nanoscale has emerged as a fertile ground for new physical phenomena 
we have analyzed in this paper FE nanoparticles based on a microscopic model. To this aim 
the Ising model in a transverse field is modified in such a manner that the effects of surfaces and defects 
are taken into account. Eventually, these properties are manifested within our model by introducing different 
microscopic coupling parameters for the constituents of the FE material. Due to the broken translational  
invariance the Green's function technique has to be formulated in real space. As the result 
of the equation of motion method for the Green's function we find the polarization of the nanoparticle 
depending on an external electric field, the temperature, the defect position and concentration as well as 
the size of the particle. In particular, we are interested in the coercive field $E_c$, which is very sensitive 
to the interaction strength at the surface $J_s$ and the presence of defects manifested by 
a microscopic coupling $J_d$. In case of a defect coupling strength that is lower than the bulk one, $J_d < J_b$, 
the coercive field, the remanent polarization $\sigma_r$ and the Curie temperature $T_c$ are reduced in 
comparison to the case without defects. This theoretical finding gives an explanation of the experimentally 
observations in FE small particles by substituting of doping ions, 
such as La in PTO [11], PZT [12,13] nanopowders. Contrary we get in the opposite case with 
$J_d > J_b$ (for example when the impurities have a larger radius compared with the 
constituent ions), that the coercive field, the remanent polarization and the critical temperature  
are larger than the case without defects. This realization is in accordance with the experimental 
data showing an increase of $T_c$, $E_c$ and $\sigma_r$ by the substitution of doping ions, such as 
Bi in SBT [10] or by increasing Ba contents in PLZT [14] ceramics. The dependence on the particle 
size is also discussed. Our results give strong indications that microscopic details of the interaction 
within the FE nanoparticles are essential for the macroscopic behavior of such a quantity as 
the hysteresis loop, whose shape seems to be important for practical applications of FE. 
In this paper we have demonstrated that one of the standard model for describing ferroelectric 
systems, namely the modified Ising model in a transverse field, is able to give explanation 
for experimental observations.   

\begin{acknowledgments} 
\noindent One of us (J. M. W.) is grateful to the Cluster of Excellence in Halle  
for financial support. This work is further supported by the SFB 418. Further we acknowledge 
discussions with Marin Alexe and Dietrich Hesse, MPI Halle. 
\end{acknowledgments}

\end{document}